\documentclass[reprint,showpacs,preprintnumbers,amsmath,amssymb,prb]{revtex4-1}
\usepackage[pdftex]{graphicx}
\usepackage{dcolumn}         
\usepackage{bm}              
\usepackage{verbatim}        
\usepackage{longtable}
\begin{document}
\title{Polymerization of sodium-doped liquid nitrogen under pressure}

\author{ M. M. E. Cormier$^{1,2}$,  S. A. Bonev$^{2}$}
\email[Electronic address:]{bonev@llnl.gov}
\affiliation{ $^1$Department of Physics, Dalhousie University, 
              Halifax, Nova Scotia B3H 3J5, Canada.\\
              $^2$ Lawrence Livermore National Laboratory, 
              Livermore, California 94550, USA.\\ }

\begin{abstract}
First-principles molecular dynamics (FPMD) simulations are performed on
6 and 12\% Na in dense liquid N. A detailed description of structural and electronic
properties leading to an understanding of the effect of Na-doping 
on the polymerization phase transition of N is presented. Compression of the mixtures from 5 to 90~GPa shows three distinct 
regions of characteristic local order separated by pressures near 
30 and 65~GPa. Computation of Gibbs free energies of mixing 
shows that these mixtures are thermodynamically stable beyond 
20 and 15~GPa for 6 and 12 \% Na respectively. 
\end{abstract}
\date{\today}
\maketitle

%
\section{Introduction \label{intro}}
%

High pressure ($P$) and temperature ($T$) conditions offer unique ways for
the synthesis of materials with novel mechanic, optical, and
electronic properties. In molecular solids and liquids, the breaking of
intra-molecular bonds upon compression may lead to the formation of
extended (polymeric), covalently bonded structures. The interest in
such behavior is partially motivated by the potential discovery of novel energetic
materials. Nitrogen, in particular, has become a subject of intense 
research in recent years. A polymeric structure called cubic-gauche
(cg-N), which was predicted~\cite{mailhiot_PRB_1992} among 
suggestions of possible stable monatomic solid
phases~\cite{mcmahan_PRL_1985,martin_PRB_1986},  
was eventually synthesized by Eremets {\it et al.} \cite{eremets_nmat_2004} 
at $P$ and $T$ above 110~GPa and 2000~K. 
If polymeric N is recovered
to ambient $P$ and $T$, it would be an energetic material 
with an energy capacity over five times
greater than current energetic
materials~\cite{eremets_nmat_2004}. However, attempts to quench cg-N
to ambient conditions~\cite{barbee_PRB_1993,zhang_PRB_2006,chen_PRB_2008} 
have been unsuccessful to date.

A possible route to metastable N is to tune the polymerization transition by
combining (doping) N with small
amounts of other elements. 
Indeed, in order to discover energetic materials with optimal
properties, it is essential to explore bonding and structural
properties as a function of all 
three fundamental variables -- pressure, temperature and chemical composition. 
Here optimal properties include metastability at ambient conditions,
polymerization transition at relatively low pressure, and high energy
capacity. The last requirement suggests a focus on N-rich systems with
small amounts of dopants. The
fact that molecular-to-polymeric transitions are
usually observed at high $T$ also means that it 
is critical to understand the evolution with pressure 
of bonding and structural properties of molecular compounds at 
elevated temperatures. 

Achieving these goals is challenging for a number of reasons:
(i) the extreme complexity of the phase diagrams of s-p-valent
materials, including numerous phases and extended regions of
metastability; (ii) limitations of {\it in situ}
measurements at high-$P$ and $T$; and (iii) limitations 
of theoretical techniques for
structure searches at high $T$ and large crystalline unit cells (small
dopant concentrations). Perhaps for these reasons, studies to day
\cite{Steele:CPL2015,Steele:JCP2015,Wang:JCP2014,Wang:JCP2013,Eremets:JCP2004} 
have focused mostly on compositionally simple systems, and even their
finte-$T$ phase diagrams are not entirely known yet.

Here we have taken an alternative approach to study
polymerization in N-rich systems. 
Investigations of liquids provide a direct way of studying the
evolution of bonding properties with density at finite temperature,
while avoiding many of the difficulties associated with global 
phase space searches of
crystalline stability. The relevance of these findings for the
problem of crystalline stability is grounded on the 
established parallels between solid and liquid structures under
pressure~\cite{Tamblyn:2008fh,Tamblyn:2010kp,boates_PRL_2009,boates_PRB_2011}. 
The case of N is particularly favorable because of the 
existence of sharp and well-defined structural 
changes in the dense liquid~\cite{boates_PRL_2009}.

In this paper, we report results on the polymerization of 
liquid N doped with Na. 
First principles molecular dynamics (FPMD) simulations are used to collect
statistical information about the liquid as a function of 
pressure for different Na concentrations. 
Their structural, electronic, and thermodynamic properties
are characterized, based on which distinct transition stages are
identified. Our results show that only a small concentration of Na (a
few \%)  is required to induce polymerization at 
sufficiently lower pressure compared to the pure N system.

%
\section{Computational Methods \label{comp}}
%

FPMD simulations of liquid N mixed with 6, 12, 25, and 50\% Na, as
well as pure N and pure Na,  have been performed 
for pressures from 5 to 100~GPa along their respective 2000~K isotherms using finite-temperature 
density functional theory (DFT)~\cite{kohn_PR_1965} within the Perdew-Burke-Ernzerhof 
generalized gradient approximation (PBE-GGA)~\cite{perdew_PRL_1996} as implemented 
in  the Vienna {\it ab initio} simulation package (VASP)
~\cite{vasp}.  The FPMD simulations were carried out in the canonical
(constant number of particles $N$, volume $V$, and  temperature $T$) 
ensemble using Born-Oppenheimer dynamics and a Nos$\acute{e}$-Hoover thermostat. 
Supercells with periodic boundary conditions
of 128 atoms were used for the Na-N mixtures and for pure liquid N, 
while 64 atom supercells were used for pure liquid Na. 
The supercells were initially randomly populated with N$_2$ molecules and the 
appropriate concentration of isolated Na atoms while enforcing a minimum 
separation between molecules and Na atoms. The Na concentration is
measured as the ratio of the Na atoms to the total number of atoms in
the system. 
Before compression, the liquids were confirmed to be molecular N$_2$ with isolated Na atoms.  

To obtain well-converged pressures (within $\sim 0.1$~GPa) and energies ($\sim$~meV/atom),
the Brillouin zone was efficiently sampled using a 
single $\mathbf{k}-$point at $(\frac{1}{4},\frac{1}{4},\frac{1}{4})$ in
the Brillouin zone. Convergence tests with 
up to $3\times3\times3$ uniform $\mathbf{k}-$point meshes were carried out over
a wide range of pressures and temperatures. 
Using a 0.5~fs ionic time step, each FPMD trajectory was 
first equilibrated within 2-3~ps and continued for additional 5~ps  
from which structural, electronic, and thermodynamic properties were calculated. 
Nine- and five-electron projector augmented wave (PAW) pseudopotentials (PP) 
with 2.2 and 1.6~Bohr core radii were used for all Na and N 
calculations respectively, using a 600~eV plane-wave cutoff energy. 

For accurate determination of free energies of mixing, the PBE-GGA
energies and pressures were corrected using the HSE06 hybrid functional 
of Heyd, Scurseria, and Ernzerhof~\cite{heyd_JCP_2003}. 
This was done on isolated atomic configurations taken from the FPMD
trajectories. The resulting corrections were averaged 
over the sampled snapshots. We found that five well-separated configurations from a given
trajectory are sufficient to converge the average corrections for all
pressures considered. The fluctuations in the energy corrections were
negligible, indicating that the PBE-GGA ensemble is sufficient. 


\section{Results \label{results}}

Upon compression, the Na-N mixtures exhibit three distinct 
regions with characteristic local structural order 
emerging at concentration-dependent  pressures. 
In Section~\ref{structure} we describe in detail the case of 12\% Na, 
noting that this concentration is qualitatively similar to 6 \% Na, but
that there are more significant differences at higher concentrations.
These differences will be described in a follow-up article. 
The structural analysis allows for informative decomposition of the
electronic properties of the liquids among characteristic
clusters. The electronic properties described in Section~\ref{electronic}
clearly elucidate the effect of Na-doping on the evolution of bonding
properties of N as a function of compression. Finally, 
thermodynamic stability as a function of concentration is 
analyzed for the entire pressure range considered, and concentration-dependent equations of state 
are presented in Section~\ref{free_energy}.

\subsection {Structural Properties \label{structure}}

For the case of 12\% Na, structural transitions
occur near 30 and 65 GPa. These pressures mark the 
appearance or disappearance of some characteristic features in the 
local order, while compression within each region yields gradual 
progression of structural properties with no qualitative changes. 

The initial step in the structural analysis is to calculate the
distribution of nearest neighbor (n.n.) species around each atom in
the supercell for every time step in the simulation trajectories. These data are used to identify
distinctive atomic arrangements within the liquid -- their composition,
geometry, and temporal stability.  

\begin{figure}[tb]
\includegraphics[width=\columnwidth,clip]{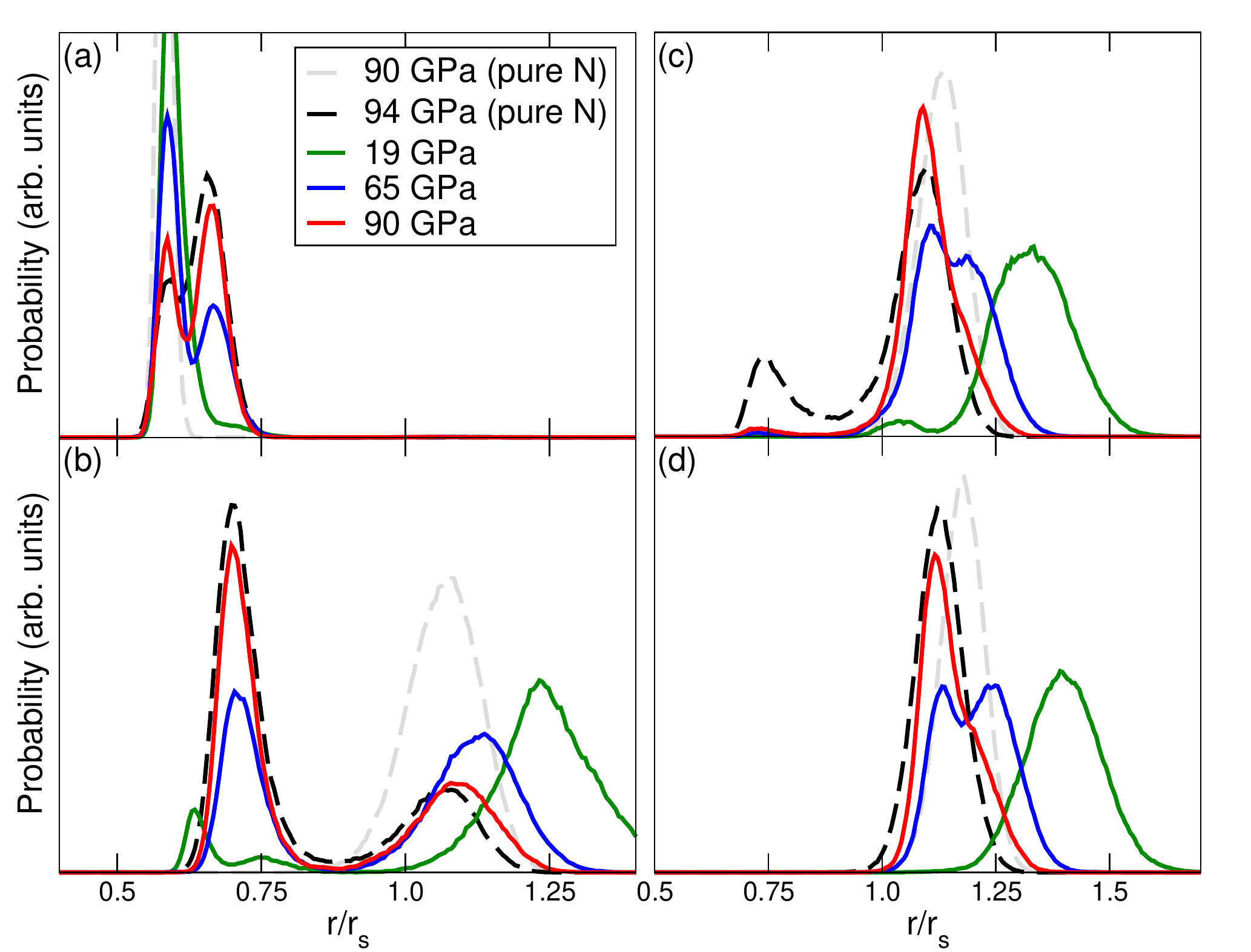}
\caption{ \label{nn_resc} Density-rescaled N-N nearest-neighbor 
          distance distributions for the (a) 1$^\mathrm{st}$, (b)
          2$^\mathrm{nd}$, (c) 3$^\mathrm{rd}$, and (d)
          4$^\mathrm{th}$ n.n. Results are shown for 12\% Na mixture,
          and for pure N at several pressure as indicated in the legend.}
\end{figure}

We found that the transitons are clearly marked by distinct features
in the 1$^\mathrm{st}$ to 4$^\mathrm{th}$ N-N
n.n. distance distributions.  Fig.~\ref{nn_resc} shows a histogram of
N-N neighbor distances, rescaled by the density parameter, $r_s$, in order
to facilitate comparison between structural properties at different
pressures. Here $r_s$ is defined by $4/3 \pi r_s^3 = V/N$; notice that
$N$ is the number of ions, not electrons.  An appropriate cut-off
radius, $R_c$ was 
determined as the first minimum of the N-N pair correlation function, from which
coordination fractions were determined as ensemble averages.
Using this information, we define a molecule 
as two atoms that are mutual nearest-neighbors and both are  
singly-coordinated (within the cut-off radius). An $n$-member ring ($R_n$)
is identified as a closed sequence 
of $n$  2-coordinated atoms . An $n$-member 
chain ($C_n$) is located by searching for a sequence of $(n-2)$ 2-coordinated atoms
terminating on each end with atoms which are $c$-coordinated, where $c\ne 2$.
Notice that with these definitions, if a cluster is formed by attaching
a $R_n$ to a $C_m$, the cluster will be identified as $C_m$.

\begin{figure}[tb]
\includegraphics[width=\columnwidth, clip]{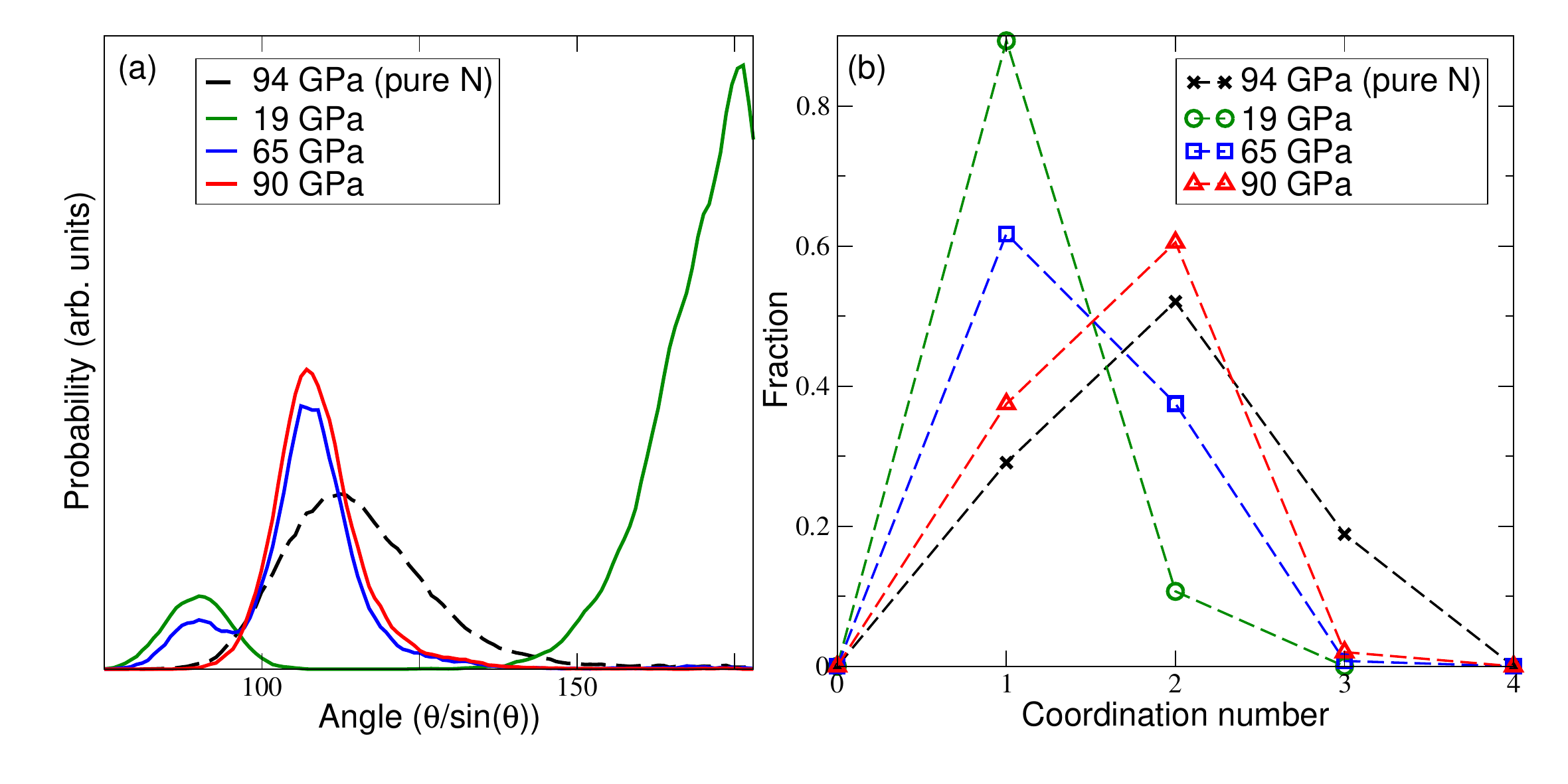}
\caption{ \label{ang_and_crd} (a) N-N-N angle distributions and (b) N-N
  coordination numbers for 12\% Na concentration in liquid N.
  The angle distributions are calculated by taking into account only the
  first two n.n. within a distance less than the appropriate cut-off
  radius  for the given pressure (see text).  The N-N coordination numbers are calculated
  based the ensemble-averaged number of N atoms found within the cut-off radius.}
\end{figure}

First, the region below 30~GPa is characterized by a broadened peak 
with a shoulder in the 1$^\mathrm{st}$ n.n. distance 
distribution and two small peaks in 
the 2$^\mathrm{nd}$ n.n.  The peak in the 1$^\mathrm{st}$  n.n. is due
to N$_2$ molecules, the broadening comes from the first neighbor
in N$_3$ chains (azide), and the shoulder corresponds to N$_4$ rings. 
The N$_3$ chains yield the first peak in the  2$^\mathrm{nd}$
n.n. distance distribution, while the second peak is due to the
formation N$_4$ rings. The N$_2$ molecules have an 
equilibrium bond distance very close to 1.1~{\AA} in agreement with 
the chemical bond distance of isolated N$_2$ molecules.
The N$_3$ chains have bond distances between 1.15 and 1.25~{\AA}, and the N-N-N angle 
is nearly flat with a peak near 180 degrees (see Fig.~\ref{ang_and_crd}) in accordance with  
double-bonded azide chains.
The 4-member N rings
are bonded with distances between 1.25 and 1.35~{\AA} as is the case in the tetra-nitrogen
molecule, corresponding to double and single bonds, and
are very close to square with a peak in the N-N-N angle near
90~degrees. Thus, at pressures up to 30~GPa, even though there is a
small fraction of borken N$_2$ molecules, the species that appear in
the liquid are known from ambient-$P$ chemistry.

\begin{figure}[tb]
\includegraphics[width=\columnwidth, clip]{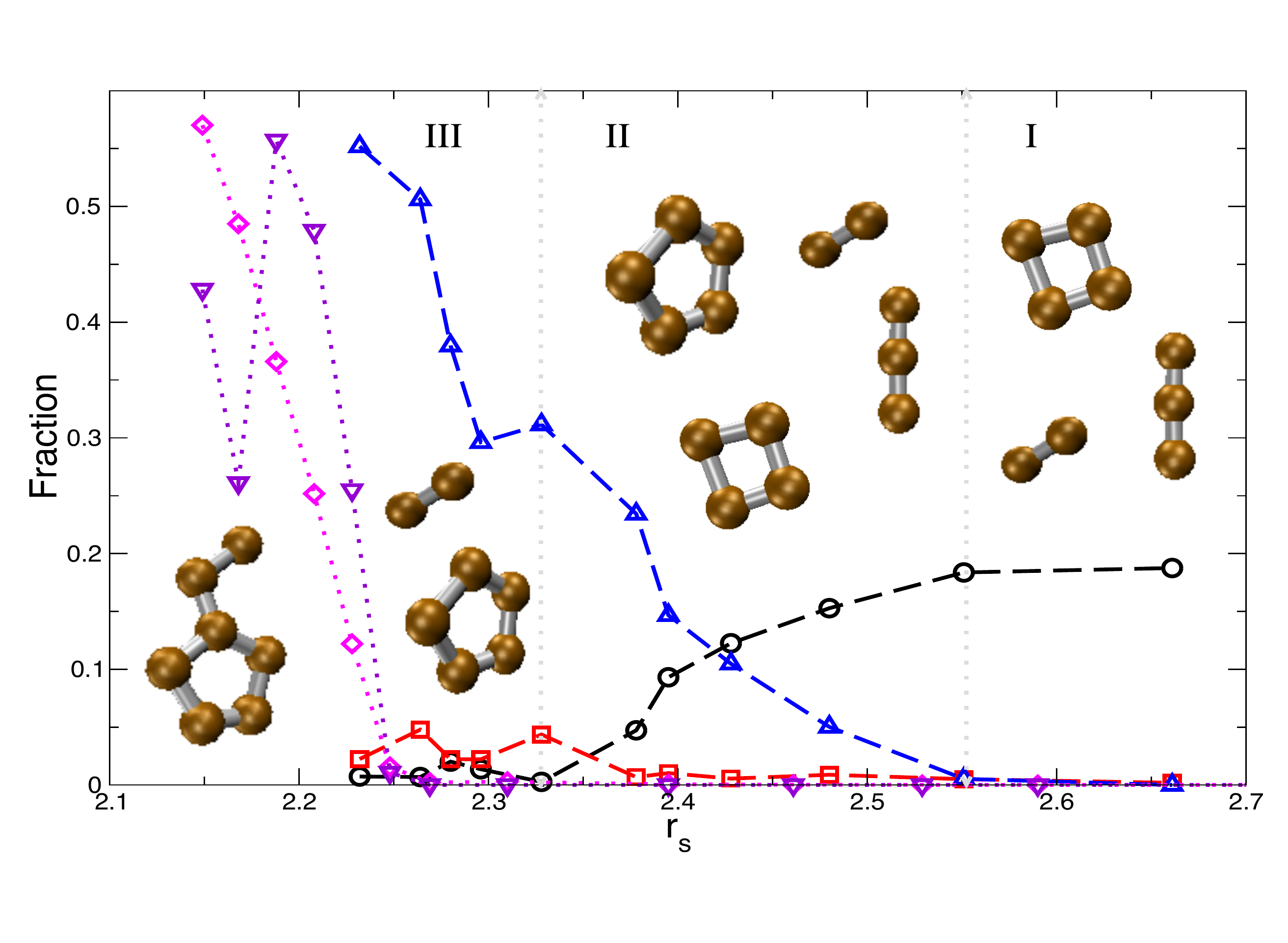}
\caption{ \label{fractions} The fraction of the total number of 
          atoms in 3-member chains for 12 \% Na (black-dashed circles), 
          4-8-member chains for 12 \% Na (red-dashed squares) and 
          pure N (magenta-dotted diamonds), 
          and 5-member rings for 12 \% Na (blue-dashed up-triangles) and 
          pure N (violet-dotted down-triangles). 
          The illustrations show characterisitic clusters present in 
          each of the three regions.} 
\end{figure}

Compression beyond 30~GPa causes N$_3$ chains to bond with  
N$_2$ molecules forming 5-member N rings. 
This is marked by the splitting of the 1$^\mathrm{st}$ n.n. peak,
the heightening of the first peak in the 2$^\mathrm{nd}$  n.n., and the
splitting of the peak in the 3$^\mathrm{rd}$ and 4$^\mathrm{th}$ 
n.n. The emergence of a peak at about 108 degrees in the angle
distribution also marks the formation of 5-member N rings. 
The fraction of these rings 
increases with pressure while the fraction of N$_3$ chains decreases
proportionally to the fraction of N$_2$ molecules (see Fig.~\ref{fractions} 
The 4-member N rings persist in this pressure region. The bond distances in the 
5-member N rings are between 1.25 and 1.35~{\AA} consistent with the
lengths of double and single N-N bonds.  

Further compression past 65 GPa yields N chains
longer than 3 while the fraction of 5-member N rings continues to increase.
The fraction of N$_3$ chains becomes negligible by 65~GPa as is 
marked by the disappearance of the first small peak in the 
2nd nearest-neighbour and the disapperance of the peak at about 
170 degrees in the angle distribution. 
The fraction of N chains longer than 3 in this region becomes non-zero and 
N chains of any length possess both the same N-N bond lengths
between 1.25-1.35 {\AA}, and N-N-N angle distribution has a peak at 108 degrees 
as the 5-member N rings. Thus, in this third region,  
3-member N chains, $n$-member N chains, with $n>3$, and 
5-member N rings cannot be distinguished simply by observing the 
neighbour distributions and N-N-N angles; coordination analysis
must be used to explicitely search for each cluster type. 
This analysis also shows that N chains in this region emerge 
as tails of 5-member N rings where one terminal N is 3-coordinated
as a member of both a ring and a chain, with the other 
terminal N singly coordinated. The complete disappearance of 
N$_4$ rings does not occur until 75 GPa. 

\subsection {Electronic Density of States \label{electronic}}

\begin{figure} [tb]
 \hspace*{-5mm}
\includegraphics[width=\columnwidth, clip]{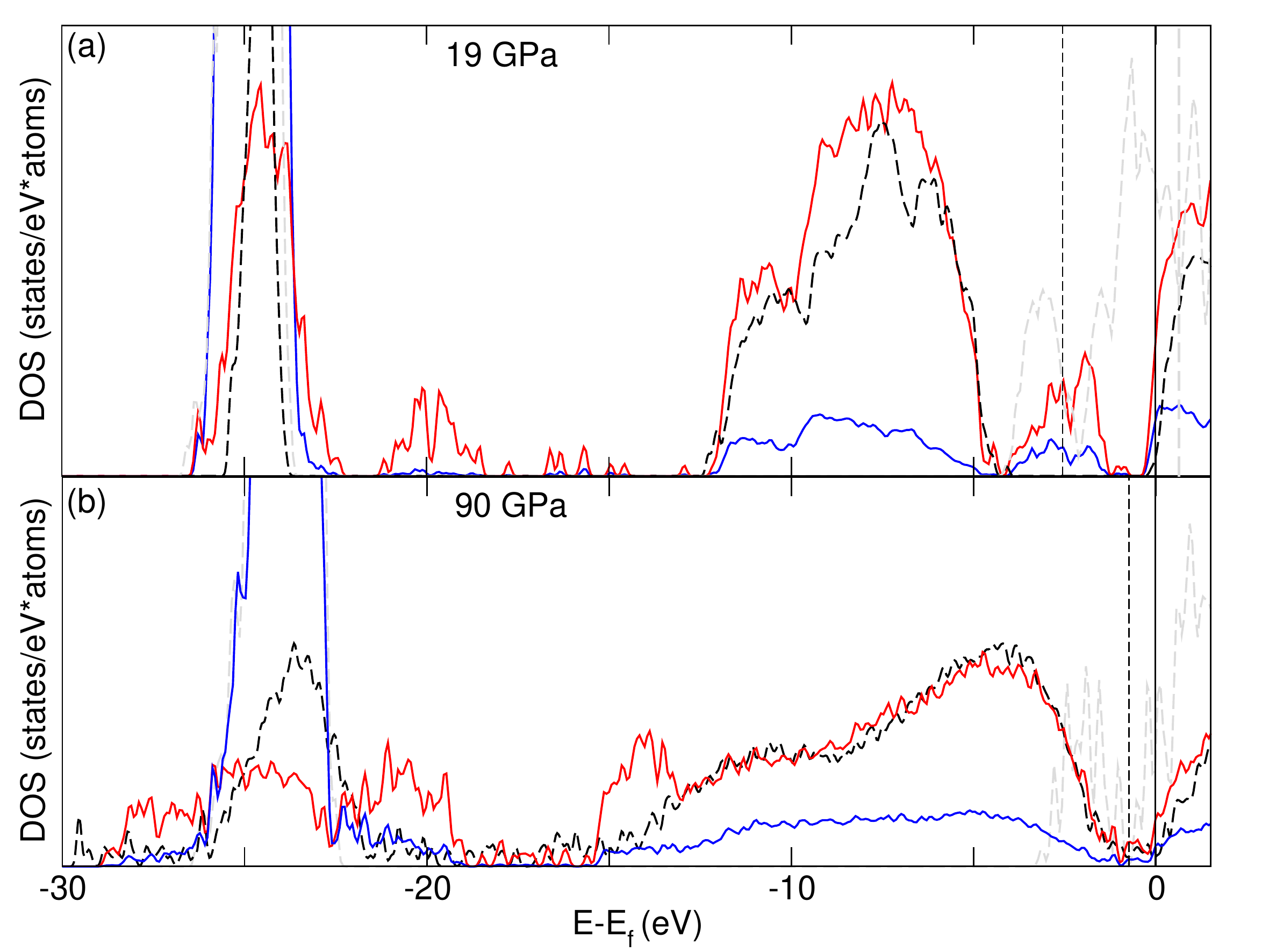}
\caption{ \label{DOS_19-90} Site-projected EDOS for 12\% Na in 
         liquid N showing the N (red) and Na (blue) 
         projections along with
         pure N (black dashed) and pure Na (grey dashed) for (a) 19~GPa 
         and (b) 90~GPa. The pure system DOS were shifted relative their
         Fermi levels to show similarities in profiles with the mixtures. 
         The repesctive Fermi levels are shown by vertical lines
         for the mixtures (black-solid), pure N (black-dashed), and
         pure Na (grey-dashed).}
\end{figure}

\begin{figure} [tb]
 \hspace*{-5mm}
\includegraphics[width=\columnwidth, clip]{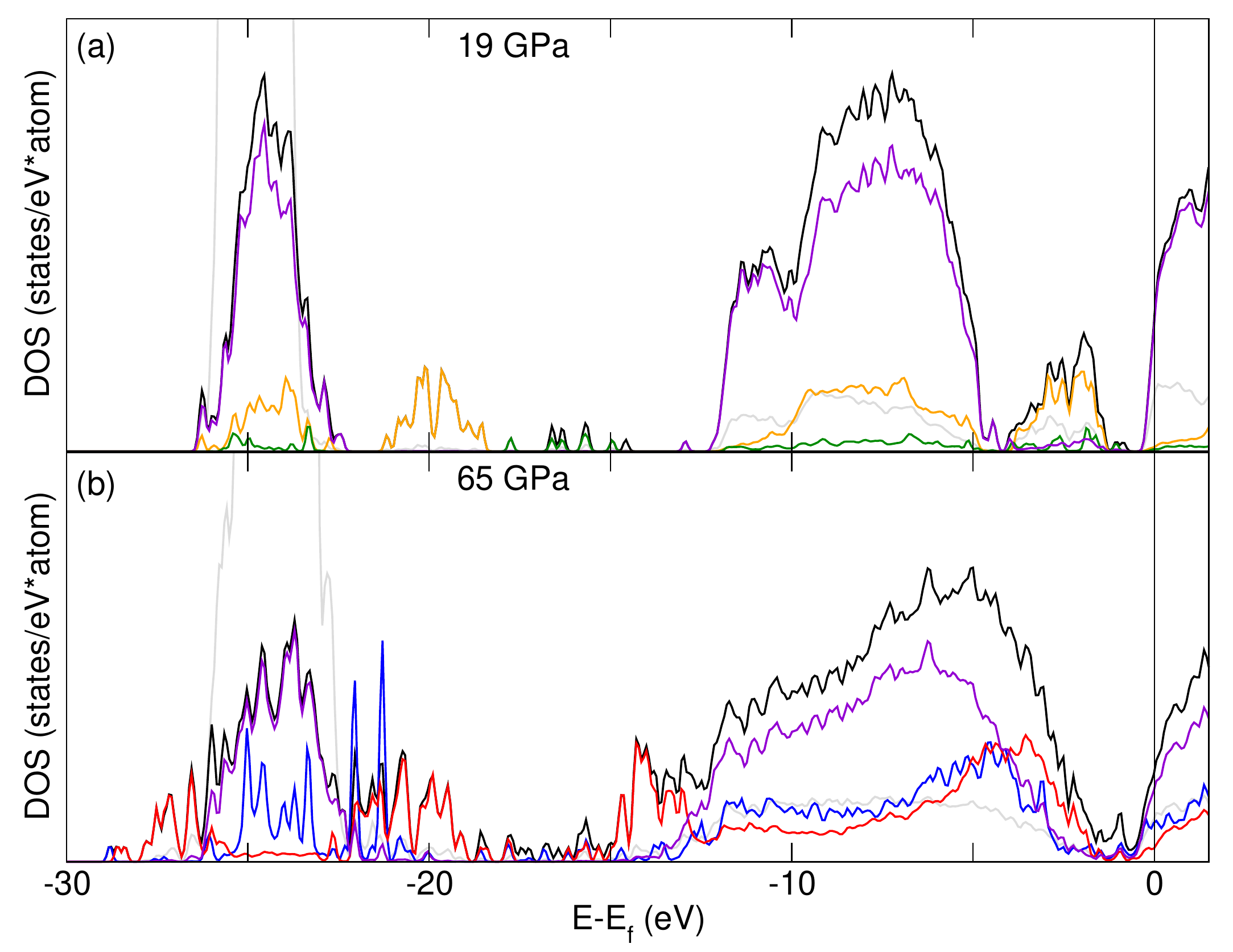}
\caption{ \label{ClusterDOS} Cluster decomposition of the electronic
         density of states for N$_3$ chains (orange), 
         N$_2$ molecules (violet), N$_4$ rings (green),
         5-member N-rings (red), and $n$-member N-chains, $n\ge 3$ (blue).
         The total projections are also shown for N (black) and 
         Na (grey).} 
\end{figure}

The electronic density of states (EDOS) are obtained from the charge density 
of atomic configurations taken from the FPMD trajectories. For liquids, the ensembles are sampled 
and the results are averaged. In our case, we found that taking five 
snapshots is sufficient to obtain converged EDOS averages. For each static 
calculation, the Brillouin zone was sampled efficiently with 4 $\mathbf{k}$-points at 
$(\frac{1}{4},\frac{1}{4},\frac{1}{4})$, 
$(\frac{1}{2},\frac{1}{4},\frac{1}{4})$,
$(\frac{1}{2},\frac{1}{2},\frac{1}{4})$, and
$(\frac{1}{2},\frac{1}{2},\frac{1}{2})$,
which were given weights 1, 3, 3, and 1, respectively. 
This sampling approach produces better results than a uniform 
$3\times3\times3$ mesh, and completely reproduces results obtained
with uniform $4\times4\times4$ and 
$6\times6\times6$ $\mathbf{k}$-point grids.
The EDOS are projected on the N and Na species (Fig.~\ref{DOS_19-90})
and further decomposed based on the types of clusters that the atoms are
part of (Fig.~\ref{ClusterDOS}).  
Site-projections were obtained by choosing a cut-off radius 
of half the 1$^\mathrm{st}$ coordination shell radius. As a consequence, 
small overlaps were incurred, but features from higher energy states that 
were otherwise missed became apparent.  

At 19~GPa (region I of the mixed system) pure N is still an insulator with a
band gap of about 5~eV. The Na atoms contribute electronic
states in this band gap region. As a result, there is a small amount of
occupied states at the Femi level of the Na-doped system - partial
occupancy of the anti-bonding orbitals of
the N$_2$ molecules. Furthermore, additional EDOS appear below
the Fermi level, which belong mostly to N$_3$ chains. These are formed
with extra electronic charge transferred from the Na atoms.  
Meanwhile, the valence 
states due to N$_2$ molecules remain very similar to those 
in pure liquid N. The molecular states around -30~eV are broadened 
compared to the pure case and the appearance of states around 
-20~eV are strictly due to N$_3$ chains. 

With increasing density, 
the edge of the molecular states is broadened towards the Fermi level
until it overlaps with the N$_3$ states in the gap.  At this point, around
30 GPa, the formation of 5-member N rings, by combining N$_2$ molecules
with N$_3$ chains, is possible. As more rings form, 
the states within the gap are completely overtaken. Notice that states owing 
to $n-$member chains ($n \ge 3$) are nearly identical to 5-member
rings, except for the lower energy stabilizing states 
around -14 eV. 

Compression beyond 65~GPa into region III yields no qualitative
changes in the electronic structure. The fraction of N$_2$ molecules
decreases and so the total number of states due to various 
clusters changes, but there is no qualitative change in profile of the
EDOS.

We see that the mechanism of polymerization here is different than in pure
N where the molecules remain stable until the electronic band gap
closes at around 90~GPa and the liquid becomes metallic. Small
amounts of Na are sufficient to change significantly 
the electronic structure of the
liquid by facilitating the formation of N$_3$ clusters with states
below the Fermi level. The latter become precursors for the breaking of
the already weakened N$_2$ molecules and the formation of poly-N at
pressure much lower than 90~GPa.

\subsection {Free Energy of Mixing and Equation of State \label{free_energy}}

Here we examine the thermodynamic stability of the liquids by computing
their Gibbs free energies. Pressures, temperatures 
and energies were calculated from ensemble (time) averages, while 
entropies were determined following the method outlined in
Ref.~\onlinecite{Teweldeberhan:2011fd}.
We then calculate the Gibbs free energies of mixing of the liquids as
 $\Delta G_{mix} = G_{(N-Na)_{i}} - x_{i}G_{N} - (1-x_{i})G_{Na}$,
where $x_{i}$ is the concentration of N 
in mixture $(N-Na)_{i}$ and each $G_{X}$ term on the right hand side 
is the Gibbs free energy per atom of liquid $X$.

\begin{figure} [tb]
\includegraphics[width=\columnwidth, clip]{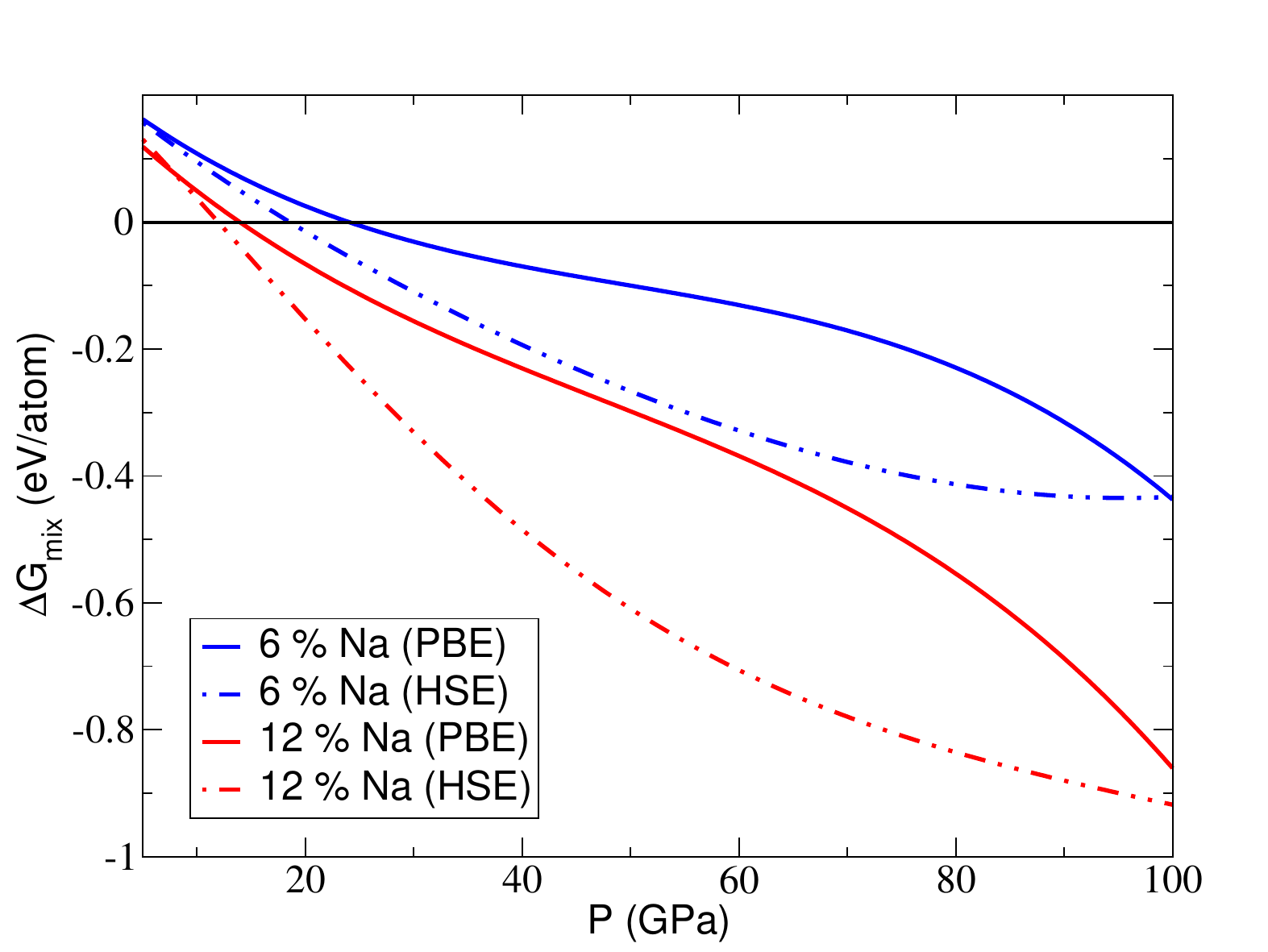}
\caption{ \label{Gmix} Gibbs free energies of mixing as a function
          of pressure with (dotted-dashed) and without (solid) HSE 
          corrections for 6 and 12 \% Na in liquid N.} 
\end{figure}

The results in Fig.~\ref{Gmix} show that the mixtures are  thermodynamically 
stable with respect to the pure systems beyond a maximum of $\sim 23$~GPa. 
Hybrid functional corrections are shown to be largest in the 
transition region, 
but yield corrections in favor of the mixtures. 

\begin{figure} [tb]
\includegraphics[width=\columnwidth, clip]{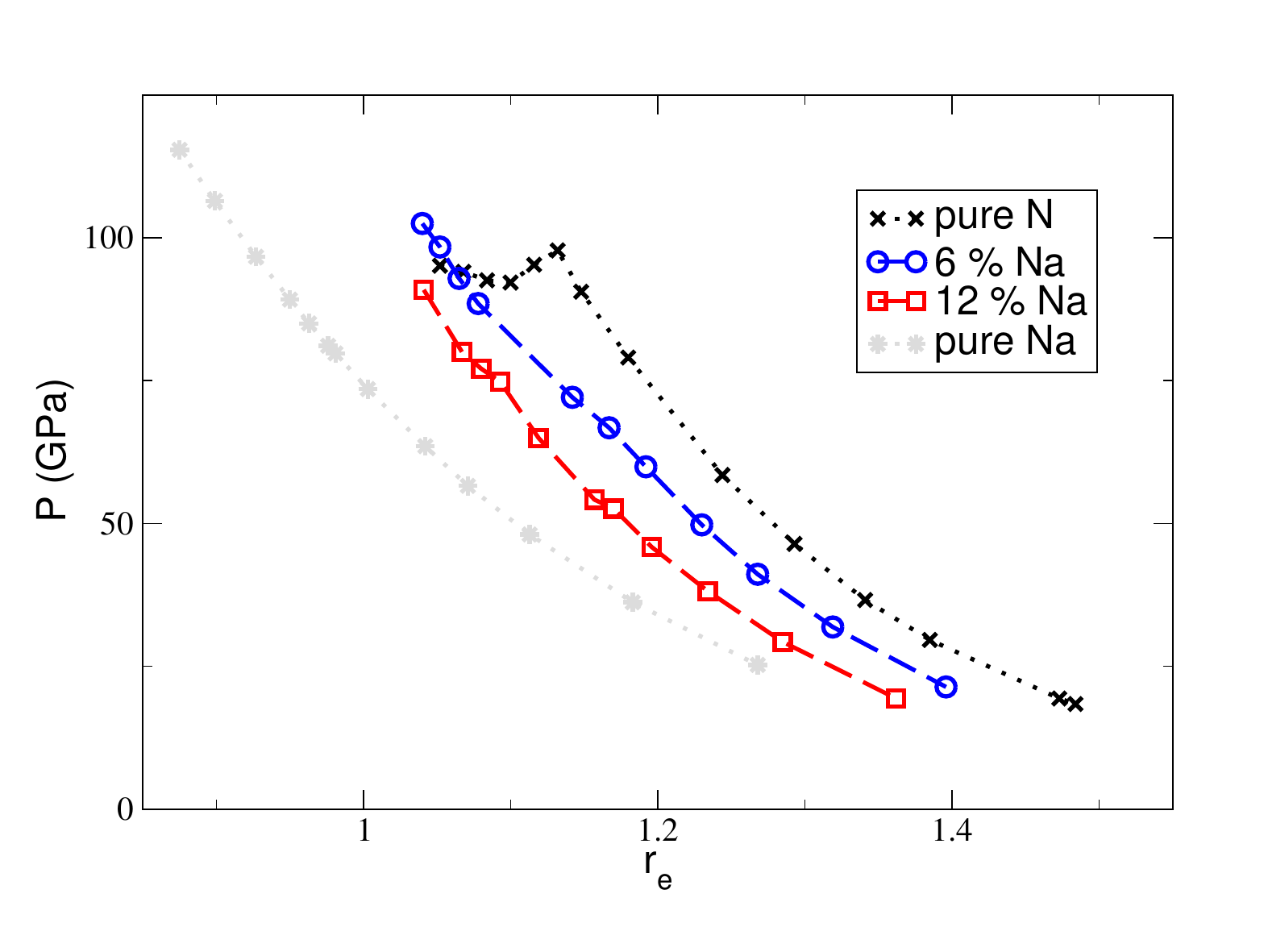}
\caption{ \label{EOS} Equations of state. 
         The pressure is plotted against the 
         valence electron density, $r_{e}$ (see text for details).} 
\end{figure}

The equations of state for pure liquid N,  N mixed with 6, and 12
\% Na, and pure liquid Na are shown in Fig. ~\ref{EOS}
The pressure is plotted against the valence electron density 
parameter, $r_{e}$, defined as $V_{e}/N_{e} = (4/3)\pi (r_{e}a_{0})^3$.
Here $V_{e} = V - V_{ion\ core}$, where $V_{ion\ core}$ is the volume
assigned to electron cores. It is calculated by taking the core radius
as the Wigner-Seitz radius from the pseudopotentials.
$N_{e}$ is the number of 
valence electrons implemented in the pseudopotentials,  
and $a_{0}$ is the Bohr radius. At the same 
pressure, the valence electron density increases with Na-doping
concentration. Thus, the Na subsystem chemically 
compresses the N subsystem, catalyzing the formation of 
5-member N rings and $n$-member chains, $n\geq3$. 

\section {Conclusion \label{conclusion}}

We have characterized the structural, electronic, and thermodynamic
properties of Na-doped liquid N as a function of pressure. The
polymerization transition develops in three stages and commences at
much lower pressures compared the the pure N system even 
with small Na concentrations. We have 
elucidated the mechanism by which Na promotes this transition.

In the case of pure liquid N, the valence states are 
broadened upon compression, 
where at the transition pressure the system metallizes, 
destabilizing N$_2$ molecules. At low pressure, the presence of
Na destabilizes N$_2$ bonds by donating its valence electron, 
yielding the preferred N$_3$ azide chains. 
Assisted by the chemical compression 
owing to Na ions, the 
valence states of N$_2$ molecules are broadened
upon compression, overlapping with the mid-gap
states of the N$_3$ chains, yielding the preferred N$_5$ rings. 
Further compression continues to break N$_2$ molecules.
The role of Na is then two-fold:
to provide additional electrons that destabilize molecular bonds, 
and to chemically compress the N subsystem, yielding a 2nd-order 
phase transition that does not fully metallize at high P.
This is in contrast to the first-order 
liquid-liquid phase transition of pure liquid N 
induced by metallization.

\section{Acknowledgments}

Work performed under the auspices of the US Department of Energy 
under contract No. DE-AC52-07NA27344.
Computational resources were provided by Acenet and Livermore Computing. 
                    

\bibliography{ref.bib}

\begin{thebibliography}{21}%
\makeatletter
\providecommand \@ifxundefined [1]{%
 \@ifx{#1\undefined}
}%
\providecommand \@ifnum [1]{%
 \ifnum #1\expandafter \@firstoftwo
 \else \expandafter \@secondoftwo
 \fi
}%
\providecommand \@ifx [1]{%
 \ifx #1\expandafter \@firstoftwo
 \else \expandafter \@secondoftwo
 \fi
}%
\providecommand \natexlab [1]{#1}%
\providecommand \enquote  [1]{``#1''}%
\providecommand \bibnamefont  [1]{#1}%
\providecommand \bibfnamefont [1]{#1}%
\providecommand \citenamefont [1]{#1}%
\providecommand \href@noop [0]{\@secondoftwo}%
\providecommand \href [0]{\begingroup \@sanitize@url \@href}%
\providecommand \@href[1]{\@@startlink{#1}\@@href}%
\providecommand \@@href[1]{\endgroup#1\@@endlink}%
\providecommand \@sanitize@url [0]{\catcode `\\12\catcode `\$12\catcode
  `\&12\catcode `\#12\catcode `\^12\catcode `\_12\catcode `\%12\relax}%
\providecommand \@@startlink[1]{}%
\providecommand \@@endlink[0]{}%
\providecommand \url  [0]{\begingroup\@sanitize@url \@url }%
\providecommand \@url [1]{\endgroup\@href {#1}{\urlprefix }}%
\providecommand \urlprefix  [0]{URL }%
\providecommand \Eprint [0]{\href }%
\providecommand \doibase [0]{http://dx.doi.org/}%
\providecommand \selectlanguage [0]{\@gobble}%
\providecommand \bibinfo  [0]{\@secondoftwo}%
\providecommand \bibfield  [0]{\@secondoftwo}%
\providecommand \translation [1]{[#1]}%
\providecommand \BibitemOpen [0]{}%
\providecommand \bibitemStop [0]{}%
\providecommand \bibitemNoStop [0]{.\EOS\space}%
\providecommand \EOS [0]{\spacefactor3000\relax}%
\providecommand \BibitemShut  [1]{\csname bibitem#1\endcsname}%
\let\auto@bib@innerbib\@empty
\bibitem [{\citenamefont {Mailhiot}\ \emph {et~al.}(1992)\citenamefont
  {Mailhiot}, \citenamefont {Yang},\ and\ \citenamefont
  {McMahan}}]{mailhiot_PRB_1992}%
  \BibitemOpen
  \bibfield  {author} {\bibinfo {author} {\bibfnamefont {C.}~\bibnamefont
  {Mailhiot}}, \bibinfo {author} {\bibfnamefont {L.~H.}\ \bibnamefont {Yang}},
  \ and\ \bibinfo {author} {\bibfnamefont {A.~K.}\ \bibnamefont {McMahan}},\
  }\href@noop {} {\bibfield  {journal} {\bibinfo  {journal} {Phys. Rev. B}\
  }\textbf {\bibinfo {volume} {46}},\ \bibinfo {pages} {14419} (\bibinfo {year}
  {1992})}\BibitemShut {NoStop}%
\bibitem [{\citenamefont {McMahan}\ and\ \citenamefont
  {LeSar}(1985)}]{mcmahan_PRL_1985}%
  \BibitemOpen
  \bibfield  {author} {\bibinfo {author} {\bibfnamefont {A.~K.}\ \bibnamefont
  {McMahan}}\ and\ \bibinfo {author} {\bibfnamefont {R.}~\bibnamefont
  {LeSar}},\ }\href@noop {} {\bibfield  {journal} {\bibinfo  {journal} {Phys.
  Rev. Lett.}\ }\textbf {\bibinfo {volume} {54}},\ \bibinfo {pages} {1929}
  (\bibinfo {year} {1985})}\BibitemShut {NoStop}%
\bibitem [{\citenamefont {Martin}\ and\ \citenamefont
  {Needs}(1986)}]{martin_PRB_1986}%
  \BibitemOpen
  \bibfield  {author} {\bibinfo {author} {\bibfnamefont {R.~M.}\ \bibnamefont
  {Martin}}\ and\ \bibinfo {author} {\bibfnamefont {R.~J.}\ \bibnamefont
  {Needs}},\ }\href@noop {} {\bibfield  {journal} {\bibinfo  {journal} {Phys.
  Rev. B}\ }\textbf {\bibinfo {volume} {34}},\ \bibinfo {pages} {5082}
  (\bibinfo {year} {1986})}\BibitemShut {NoStop}%
\bibitem [{\citenamefont {Eremets}\ \emph {et~al.}(2004)\citenamefont
  {Eremets}, \citenamefont {Gavriliuk}, \citenamefont {Trojan}, \citenamefont
  {Dzivenko},\ and\ \citenamefont {Boehler}}]{eremets_nmat_2004}%
  \BibitemOpen
  \bibfield  {author} {\bibinfo {author} {\bibfnamefont {M.~I.}\ \bibnamefont
  {Eremets}}, \bibinfo {author} {\bibfnamefont {A.~G.}\ \bibnamefont
  {Gavriliuk}}, \bibinfo {author} {\bibfnamefont {I.~A.}\ \bibnamefont
  {Trojan}}, \bibinfo {author} {\bibfnamefont {D.~A.}\ \bibnamefont
  {Dzivenko}}, \ and\ \bibinfo {author} {\bibfnamefont {R.}~\bibnamefont
  {Boehler}},\ }\href@noop {} {\bibfield  {journal} {\bibinfo  {journal}
  {Nature Materials}\ }\textbf {\bibinfo {volume} {3}},\ \bibinfo {pages} {558}
  (\bibinfo {year} {2004})}\BibitemShut {NoStop}%
\bibitem [{\citenamefont {Barbee{~III}}(1993)}]{barbee_PRB_1993}%
  \BibitemOpen
  \bibfield  {author} {\bibinfo {author} {\bibfnamefont {T.~W.}\ \bibnamefont
  {Barbee{~III}}},\ }\href@noop {} {\bibfield  {journal} {\bibinfo  {journal}
  {Phys. Rev. B}\ }\textbf {\bibinfo {volume} {48}},\ \bibinfo {pages} {9327}
  (\bibinfo {year} {1993})}\BibitemShut {NoStop}%
\bibitem [{\citenamefont {Zhang{~\it et al.}}(2006)}]{zhang_PRB_2006}%
  \BibitemOpen
  \bibfield  {author} {\bibinfo {author} {\bibfnamefont {T.}~\bibnamefont
  {Zhang{~\it et al.}}},\ }\href@noop {} {\bibfield  {journal} {\bibinfo
  {journal} {Phys. Rev. B}\ }\textbf {\bibinfo {volume} {73}},\ \bibinfo
  {pages} {094105} (\bibinfo {year} {2006})}\BibitemShut {NoStop}%
\bibitem [{\citenamefont {Chen}\ \emph {et~al.}(2008)\citenamefont {Chen},
  \citenamefont {Fu},\ and\ \citenamefont {Podloucky}}]{chen_PRB_2008}%
  \BibitemOpen
  \bibfield  {author} {\bibinfo {author} {\bibfnamefont {X.-Q.}\ \bibnamefont
  {Chen}}, \bibinfo {author} {\bibfnamefont {C.~L.}\ \bibnamefont {Fu}}, \ and\
  \bibinfo {author} {\bibfnamefont {R.}~\bibnamefont {Podloucky}},\ }\href@noop
  {} {\bibfield  {journal} {\bibinfo  {journal} {Phys. Rev. B}\ }\textbf
  {\bibinfo {volume} {77}},\ \bibinfo {pages} {064103} (\bibinfo {year}
  {2008})}\BibitemShut {NoStop}%
\bibitem [{\citenamefont {Steele}\ and\ \citenamefont
  {Oleynik}(2016)}]{Steele:CPL2015}%
  \BibitemOpen
  \bibfield  {author} {\bibinfo {author} {\bibfnamefont {B.~A.}\ \bibnamefont
  {Steele}}\ and\ \bibinfo {author} {\bibfnamefont {I.~I.}\ \bibnamefont
  {Oleynik}},\ }\href@noop {} {\bibfield  {journal} {\bibinfo  {journal}
  {Chemical Physics Letters}\ }\textbf {\bibinfo {volume} {643}},\ \bibinfo
  {pages} {21} (\bibinfo {year} {2016})}\BibitemShut {NoStop}%
\bibitem [{\citenamefont {Steele}\ and\ \citenamefont
  {Oleynik}(2015)}]{Steele:JCP2015}%
  \BibitemOpen
  \bibfield  {author} {\bibinfo {author} {\bibfnamefont {B.~A.}\ \bibnamefont
  {Steele}}\ and\ \bibinfo {author} {\bibfnamefont {I.~I.}\ \bibnamefont
  {Oleynik}},\ }\href@noop {} {\bibfield  {journal} {\bibinfo  {journal}
  {Journal of Chemical Physics}\ }\textbf {\bibinfo {volume} {143}},\ \bibinfo
  {pages} {234705} (\bibinfo {year} {2015})}\BibitemShut {NoStop}%
\bibitem [{\citenamefont {et~\emph{al.}}(2014)}]{Wang:JCP2014}%
  \BibitemOpen
  \bibfield  {author} {\bibinfo {author} {\bibfnamefont {X.~W.}\ \bibnamefont
  {et~\emph{al.}}},\ }\href@noop {} {\bibfield  {journal} {\bibinfo  {journal}
  {Journal of Chemical Physics}\ }\textbf {\bibinfo {volume} {141}},\ \bibinfo
  {pages} {044717} (\bibinfo {year} {2014})}\BibitemShut {NoStop}%
\bibitem [{\citenamefont {et~\emph{al.}}(2013)}]{Wang:JCP2013}%
  \BibitemOpen
  \bibfield  {author} {\bibinfo {author} {\bibfnamefont {X.~W.}\ \bibnamefont
  {et~\emph{al.}}},\ }\href@noop {} {\bibfield  {journal} {\bibinfo  {journal}
  {Journal of Chemical Physics}\ }\textbf {\bibinfo {volume} {139}},\ \bibinfo
  {pages} {164710} (\bibinfo {year} {2013})}\BibitemShut {NoStop}%
\bibitem [{\citenamefont {et~\emph{al.}}(2004)}]{Eremets:JCP2004}%
  \BibitemOpen
  \bibfield  {author} {\bibinfo {author} {\bibfnamefont {M.~I.~E.}\
  \bibnamefont {et~\emph{al.}}},\ }\href@noop {} {\bibfield  {journal}
  {\bibinfo  {journal} {Journal of Chemical Physics}\ }\textbf {\bibinfo
  {volume} {120}},\ \bibinfo {pages} {10618} (\bibinfo {year}
  {2004})}\BibitemShut {NoStop}%
\bibitem [{\citenamefont {Tamblyn}\ \emph {et~al.}(2008)\citenamefont
  {Tamblyn}, \citenamefont {Raty},\ and\ \citenamefont
  {Bonev}}]{Tamblyn:2008fh}%
  \BibitemOpen
  \bibfield  {author} {\bibinfo {author} {\bibfnamefont {I.}~\bibnamefont
  {Tamblyn}}, \bibinfo {author} {\bibfnamefont {J.-Y.}\ \bibnamefont {Raty}}, \
  and\ \bibinfo {author} {\bibfnamefont {S.~A.}\ \bibnamefont {Bonev}},\
  }\href@noop {} {\bibfield  {journal} {\bibinfo  {journal} {Physical Review
  Letters}\ }\textbf {\bibinfo {volume} {101}},\ \bibinfo {pages} {075703}
  (\bibinfo {year} {2008})}\BibitemShut {NoStop}%
\bibitem [{\citenamefont {Tamblyn}\ and\ \citenamefont
  {Bonev}(2010)}]{Tamblyn:2010kp}%
  \BibitemOpen
  \bibfield  {author} {\bibinfo {author} {\bibfnamefont {I.}~\bibnamefont
  {Tamblyn}}\ and\ \bibinfo {author} {\bibfnamefont {S.~A.}\ \bibnamefont
  {Bonev}},\ }\href@noop {} {\bibfield  {journal} {\bibinfo  {journal}
  {Physical Review Letters}\ }\textbf {\bibinfo {volume} {104}} (\bibinfo
  {year} {2010})}\BibitemShut {NoStop}%
\bibitem [{\citenamefont {Boates}\ and\ \citenamefont
  {Bonev}(2009)}]{boates_PRL_2009}%
  \BibitemOpen
  \bibfield  {author} {\bibinfo {author} {\bibfnamefont {B.}~\bibnamefont
  {Boates}}\ and\ \bibinfo {author} {\bibfnamefont {S.~A.}\ \bibnamefont
  {Bonev}},\ }\href@noop {} {\bibfield  {journal} {\bibinfo  {journal} {Phys.
  Rev. Lett.}\ }\textbf {\bibinfo {volume} {102}},\ \bibinfo {pages} {015701}
  (\bibinfo {year} {2009})}\BibitemShut {NoStop}%
\bibitem [{\citenamefont {Boates}\ and\ \citenamefont
  {Bonev}(2011)}]{boates_PRB_2011}%
  \BibitemOpen
  \bibfield  {author} {\bibinfo {author} {\bibfnamefont {B.}~\bibnamefont
  {Boates}}\ and\ \bibinfo {author} {\bibfnamefont {S.}~\bibnamefont {Bonev}},\
  }\href@noop {} {\bibfield  {journal} {\bibinfo  {journal} {Phys. Rev. B}\
  }\textbf {\bibinfo {volume} {83}},\ \bibinfo {pages} {174114} (\bibinfo
  {year} {2011})}\BibitemShut {NoStop}%
\bibitem [{\citenamefont {Kohn}\ and\ \citenamefont
  {Sham}(1965)}]{kohn_PR_1965}%
  \BibitemOpen
  \bibfield  {author} {\bibinfo {author} {\bibfnamefont {W.}~\bibnamefont
  {Kohn}}\ and\ \bibinfo {author} {\bibfnamefont {L.}~\bibnamefont {Sham}},\
  }\href@noop {} {\bibfield  {journal} {\bibinfo  {journal} {Phys. Rev.}\
  }\textbf {\bibinfo {volume} {140}},\ \bibinfo {pages} {A1133} (\bibinfo
  {year} {1965})}\BibitemShut {NoStop}%
\bibitem [{\citenamefont {Perdew}\ \emph {et~al.}(1996)\citenamefont {Perdew},
  \citenamefont {Burke},\ and\ \citenamefont {Ernzerhof}}]{perdew_PRL_1996}%
  \BibitemOpen
  \bibfield  {author} {\bibinfo {author} {\bibfnamefont {J.}~\bibnamefont
  {Perdew}}, \bibinfo {author} {\bibfnamefont {K.}~\bibnamefont {Burke}}, \
  and\ \bibinfo {author} {\bibfnamefont {M.}~\bibnamefont {Ernzerhof}},\
  }\href@noop {} {\bibfield  {journal} {\bibinfo  {journal} {Phys. Rev. Lett.}\
  }\textbf {\bibinfo {volume} {77}},\ \bibinfo {pages} {3865} (\bibinfo {year}
  {1996})}\BibitemShut {NoStop}%
\bibitem [{vas()}]{vasp}%
  \BibitemOpen
  \href@noop {} {}\bibinfo {note} {G. Kresse and J. Hafner, Phys. Rev. B {\bf
  47}, 558 (1993); Comp. Mat. Sci. {\bf 6}, 15 (1996).}\BibitemShut {Stop}%
\bibitem [{\citenamefont {Heyd}\ \emph {et~al.}(2003)\citenamefont {Heyd},
  \citenamefont {Scuseria},\ and\ \citenamefont {Ernzerhof}}]{heyd_JCP_2003}%
  \BibitemOpen
  \bibfield  {author} {\bibinfo {author} {\bibfnamefont {J.}~\bibnamefont
  {Heyd}}, \bibinfo {author} {\bibfnamefont {G.}~\bibnamefont {Scuseria}}, \
  and\ \bibinfo {author} {\bibfnamefont {M.}~\bibnamefont {Ernzerhof}},\
  }\href@noop {} {\bibfield  {journal} {\bibinfo  {journal} {J. Chem. Phys.}\
  }\textbf {\bibinfo {volume} {121}},\ \bibinfo {pages} {2780} (\bibinfo {year}
  {2003})}\BibitemShut {NoStop}%
\bibitem [{\citenamefont {Teweldeberhan}\ and\ \citenamefont
  {Bonev}(2011)}]{Teweldeberhan:2011fd}%
  \BibitemOpen
  \bibfield  {author} {\bibinfo {author} {\bibfnamefont {A.~M.}\ \bibnamefont
  {Teweldeberhan}}\ and\ \bibinfo {author} {\bibfnamefont {S.~A.}\ \bibnamefont
  {Bonev}},\ }\href@noop {} {\bibfield  {journal} {\bibinfo  {journal}
  {Physical Review B}\ }\textbf {\bibinfo {volume} {83}},\ \bibinfo {pages}
  {134120} (\bibinfo {year} {2011})}\BibitemShut {NoStop}%
\end{thebibliography}%

 \end{document}